\documentclass[twocolumn, tighten]{aastex631}

\definecolor{myblue}{HTML}{1F77B4}
\definecolor{mygreen}{HTML}{2CA02C}
\definecolor{myred}{HTML}{D62728}
\definecolor{mymagenta}{HTML}{D33682}
\definecolor{codepurple}{HTML}{C42043}

\hypersetup{
bookmarks=true,         
unicode=true,           
colorlinks=true,        
linkcolor=myblue,        
citecolor=myblue,       
urlcolor=myblue        
}

\begin{document}

\title{SN\,2023emq: a flash-ionised Ibn supernova with possible CIII emission}

\author[0000-0003-4663-4300]{M. Pursiainen}
\affiliation{DTU Space, National Space Institute, Technical University of Denmark, Elektrovej 327, 2800 Kgs. Lyngby, Denmark}
\affiliation{Department of Physics, University of Warwick, Gibbet Hill Road, Coventry, CV4 7AL, UK}

\author[0000-0002-8597-0756]{G. Leloudas}
\affiliation{DTU Space, National Space Institute, Technical University of Denmark, Elektrovej 327, 2800 Kgs. Lyngby, Denmark}

\author[0000-0001-6797-1889]{S. Schulze}
\affiliation{The Oskar Klein Centre, Department of Physics, Stockholm University, AlbaNova, SE-10691 Stockholm, Sweden}

\author[0000-0002-0326-6715]{P. Charalampopoulos}
\affiliation{Department of Physics and Astronomy, University of Turku, FI-20014 Turku, Finland}

\author[0000-0002-4269-7999]{C.~R. Angus}
\affiliation{DARK, Niels Bohr Institute, University of Copenhagen, Copenhagen, Denmark}


\author[0000-0003-0227-3451]{J.~P. Anderson}
\affiliation{European Southern Observatory, Alonso de C\'ordova 3107, Casilla 19, Santiago, Chile}
\affiliation{Millennium Institute of Astrophysics MAS, Nuncio Monsenor Sotero Sanz 100, Off. 104, Providencia, Santiago, Chile}

\author[0000-0003-0227-3451]{F. Bauer}
\affiliation{Instituto de Astrof{\'{\i}}sica, Facultad de F{\'{i}}sica, Pontificia Universidad Cat{\'{o}}lica de Chile, Campus San Joaquín, Av. Vicuña Mackenna 4860, Macul Santiago, Chile, 7820436}
\affiliation{Centro de Astroingenier{\'{\i}}a, Facultad de F{\'{i}}sica, Pontificia Universidad Cat{\'{o}}lica de Chile, Campus San Joaquín, Av. Vicuña Mackenna 4860, Macul Santiago, Chile, 7820436}
\affiliation{Millennium Institute of Astrophysics MAS, Nuncio Monsenor Sotero Sanz 100, Off. 104, Providencia, Santiago, Chile}

\author[0000-0002-1066-6098]{T.-W. Chen}
\affiliation{Technische Universit{\"a}t M{\"u}nchen, TUM School of Natural Sciences, Physik-Department, James-Franck-Stra{\ss}e 1, 85748 Garching, Germany}

\author[0000-0002-1296-6887]{L. Galbany}
\affiliation{Institute of Space Sciences (ICE, CSIC), Campus UAB, Carrer de Can Magrans, s/n, E-08193 Barcelona, Spain}
\affiliation{Institut d’Estudis Espacials de Catalunya (IEEC), E-08034 Barcelona, Spain}

\author[0000-0002-1650-1518]{M. Gromadzki}
\affiliation{Astronomical Observatory, University of Warsaw, Al. Ujazdowskie 4, 00-478 Warszawa, Poland}

\author[0000-0003-2375-2064]{ C.~P. Guti\'errez}
\affiliation{Institut d’Estudis Espacials de Catalunya, Gran Capit\`a, 2-4, Edifici Nexus, Desp. 201, E-08034 Barcelona, Spain}
\affiliation{ Institute of Space Sciences (ICE, CSIC), Campus UAB, Carrer de Can Magrans, s/n, E-08193 Barcelona, Spain}

\author[0000-0002-3968-4409]{C. Inserra}
\affiliation{Cardiff Hub for Astrophysics Research and Technology, School of Physics \& Astronomy, Cardiff University, Queens Buildings, The Parade, Cardiff, CF24 3AA, UK}

\author[0000-0002-3464-0642]{J. Lyman}
\affiliation{Department of Physics, University of Warwick, Gibbet Hill Road, Coventry, CV4 7AL, UK}

\author[0000-0003-3939-7167]{T.~E. M\"uller-Bravo}
\affiliation{Institute of Space Sciences (ICE, CSIC), Campus UAB, Carrer de Can Magrans, s/n, E-08193 Barcelona, Spain}

\author[0000-0002-2555-3192]{M. Nicholl}
\affiliation{Astrophysics Research Centre, School of Mathematics and Physics, Queens University Belfast, Belfast BT7 1NN, UK}

\author[0000-0002-8229-1731]{S.~J. Smartt}
\affiliation{Department of Physics, University of Oxford, Keble Road, Oxford, OX1 3RH, UK}
\affiliation{Astrophysics Research Centre, School of Mathematics and Physics, Queens University Belfast, Belfast BT7 1NN, UK}

\author[0000-0003-3433-1492]{L. Tartaglia}
\affiliation{INAF - Osservatorio Astronomico d'Abruzzo, via M. Maggini snc, I-64100 Teramo, Italy}

\author[0000-0002-3073-1512]{P. Wiseman}
\affiliation{School of Physics and Astronomy, University of Southampton, Southampton, SO17 1BJ, UK}

\author[0000-0002-2555-3192]{D.~R. Young}
\affiliation{Astrophysics Research Centre, School of Mathematics and Physics, Queens University Belfast, Belfast BT7 1NN, UK}






\begin{abstract}

SN\,2023emq is a fast-evolving transient initially classified as a rare Type Icn supernova (SN), interacting with a H- and He-free circumstellar medium (CSM) around maximum light. Subsequent spectroscopy revealed the unambiguous emergence of narrow He lines, confidently placing SN\,2023emq in the more common Type Ibn class. Photometrically SN\,2023emq has several uncommon properties regardless of its class, including its extreme initial decay (faster than $>90$\% of Ibn/Icn SNe) and sharp transition in the decline rate from 0.20\,mag/d to 0.07\,mag/d at $+20$\,d. The bolometric light curve can be modelled as CSM interaction with $0.32M_\odot$ of ejecta and $0.12M_\odot$ of CSM, with $0.006M_\odot$ of nickel, as expected of fast interacting SNe. Furthermore, broad-band polarimetry at $+8.7$ days ($P=0.55 \pm 0.30$\%) is consistent with spherical symmetry. A discovery of a transitional Icn/Ibn SN would be unprecedented and would give valuable insights into the nature of mass loss suffered by the progenitor just before death, but we favour an interpretation that SN\,2023emq is a type Ibn SN that exhibited flash-ionised features in the earliest spectrum, as the features are not an exact match with other SNe Icn to date. However, the feature at $5700$\,Å, in the region of \ion{C}{3} and \ion{N}{2} emission, is significantly stronger in SN\,2023emq than in the few other flash-ionised Type Ibn SNe, and if it is related to \ion{C}{3}, it possibly implies a continuum of properties between the two classes.


\end{abstract}



\section{Introduction}

The study of stellar explosions entered a new era in the wake of untargeted, high-cadence transient surveys allowing the discovery of transient phenomena that evolve on much faster timescales than typically expected of supernovae (SNe). Such rapidly evolving transients (RETs) were first discovered in archival searches in surveys such as Pan-STARRS1 Medium Deep Survey \citep{Drout2014} and Dark Energy Survey \citep{Pursiainen2018, Wiseman2020a}, resulting in large samples of spectroscopically unclassified events. Even without classifications these events expanded our understanding of stellar deaths as their light curves evolved too fast to be explained by the decay of radioactive $^{56}$Ni -- the canonical power source of many typical SNe -- and alternative power sources had to be considered. 

In recent years an increasing number of fast events have been discovered in real time allowing intense follow-up campaigns. A significant number of them appear to be interacting H-poor SNe \citep[e.g.][]{Ho2023}, either Type Ibn SNe characterised by a plethora of narrow helium emission lines \citep[e.g.][]{Foley2007, Pastorello2008}, or Type Icn with emission lines of carbon and oxygen \citep[e.g.][]{Gal-Yam2022, Perley2022}. While tens of Type Ibn SNe have been discovered over the last 15 years, so far only five Type Icn SNe have been identified -- all within the last four years. These are SN\,2019hgp \citep{Gal-Yam2022}, SN\,2019jc \citep{Pellegrino2022}, SN\,2021csp \citep{Fraser2021, Perley2022}, SN\,2021ckj \citep{Pellegrino2022,Nagao2023} and SN\,2022ann \citep{Davis2022}. Although not all Type Ibn are considered to be fast evolving \citep[e.g.][]{Inserra2019}, a significant subpopulation of them evolve on similar timescales to RETs, whilst all Icn SNe fall into this bright-and-fast parameter space. 

The existence of transitional events that change type between Ibn and IIn supernovae over time \citep[e.g.][]{Pastorello2015c, Reguitti2022}, suggests a continuum in the circumstellar material (CSM) properties of these events and consequently in the mass loss history of their progenitor stars. It may therefore be reasonable to expect that transitional events between non-hydrogen dominated CSM events (i.e. Type Icn to Ibn) also exist, although such events have not been identified to date. The discovery of such potential transitional events would be extremely valuable for our understanding of how massive stars lose their mass and create their CSM during their final stages giving insight into the chemical structure of the progenitor and the continuous or episodic nature of their mass loss.

Here we present an analysis of SN\,2023emq, a SN initially identified as a Icn SN \citep{Pellegrino2023}, but which evolved to resemble a Ibn SN $\sim10$\,d later \citep{Pursiainen2023}, possibly implying transitional nature. This paper is structured as follows: in Section \ref{sec:obs} we present the observations and the data reduction procedures, in Section \ref{sec:analysis} we present the analysis of the data and in Section \ref{sec:conclusions} we conclude our findings. The spectroscopy and photometry presented in this paper have been corrected for Milky Way extinction $E_\mathrm{B-V} = 0.105$ \citep{Schlafly2011}. Throughout the paper, we assume a flat $\Lambda$CDM cosmology with $\Omega_\mathrm{M}=0.3$ and H$_\mathrm{0}$ = 70 km s$^{-1}$ Mpc$^{-1}$ and adopt redshift $z=0.0338$.

\begin{figure}
    \centering
     \includegraphics[width=0.49\textwidth]{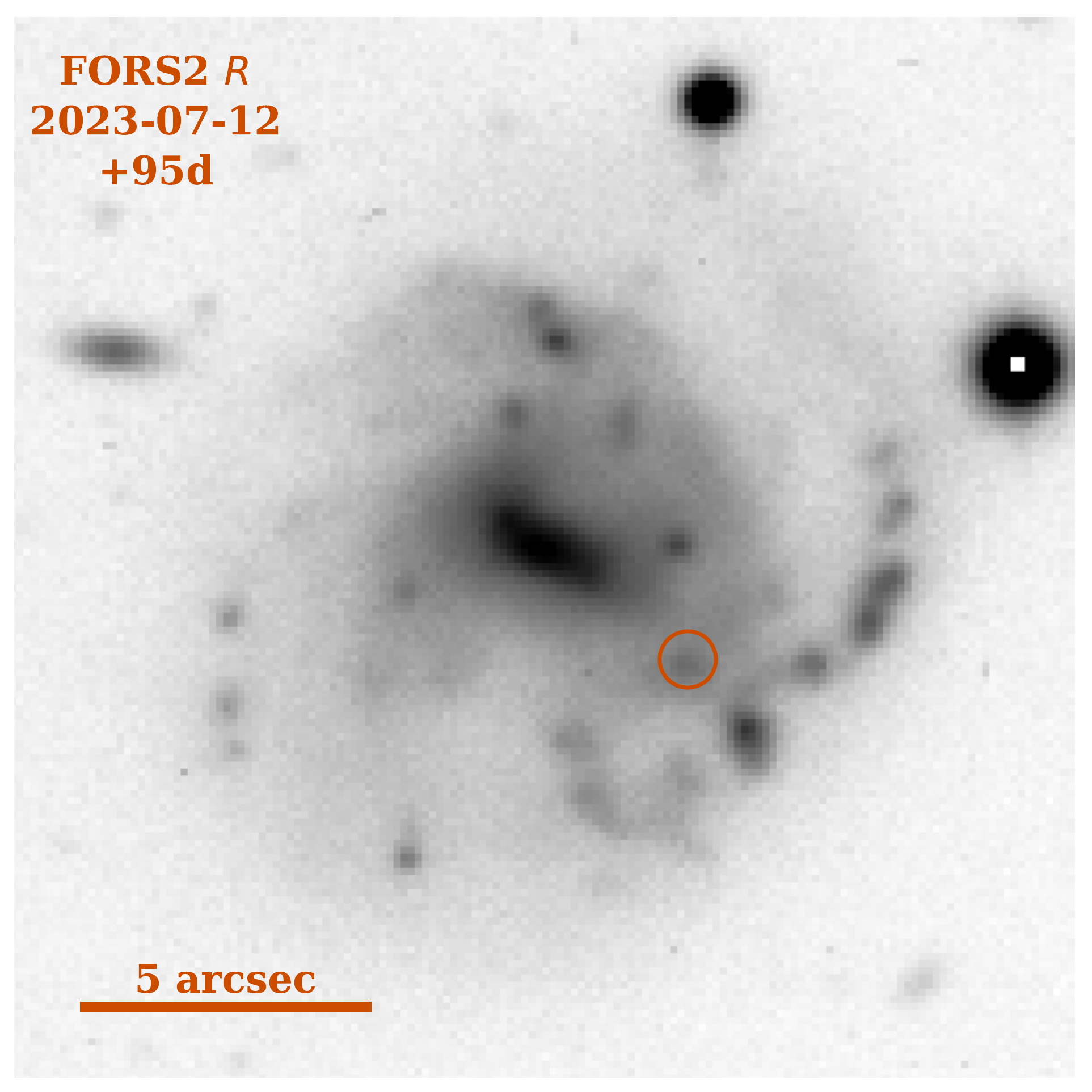}
    \caption{The host environment of SN\,2023emq in the VLT/FORS2 $R$ band observation taken at $+95$\,d (2023-07-12) when the SN was already undetectable (see Table \ref{tab:all_phot}). The location of the SN (red circle), is on top of a bright region of the host galaxy.}
    \label{fig:VLT_stamp}
\end{figure}

\begin{figure*}
    \centering
     \includegraphics[width=0.99\textwidth]{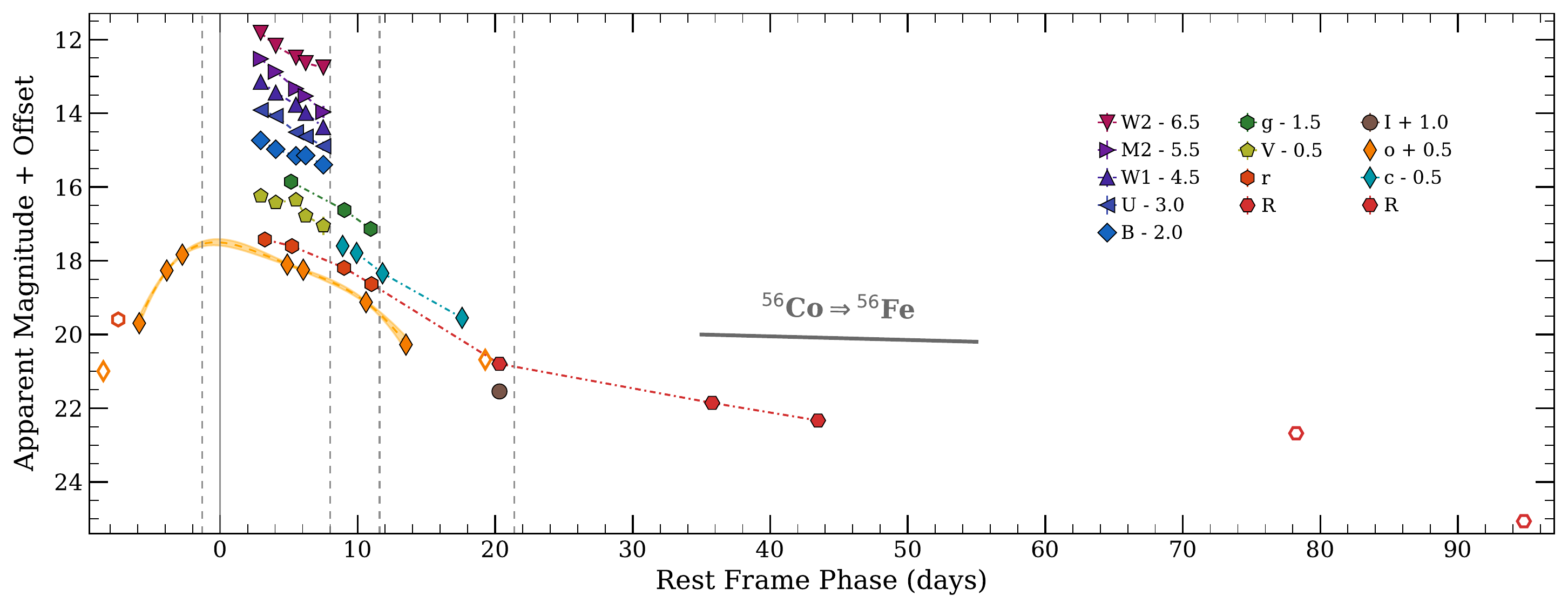}
        \begin{minipage}[b]{0.325\textwidth}
            \centering
            \includegraphics[width=\textwidth]{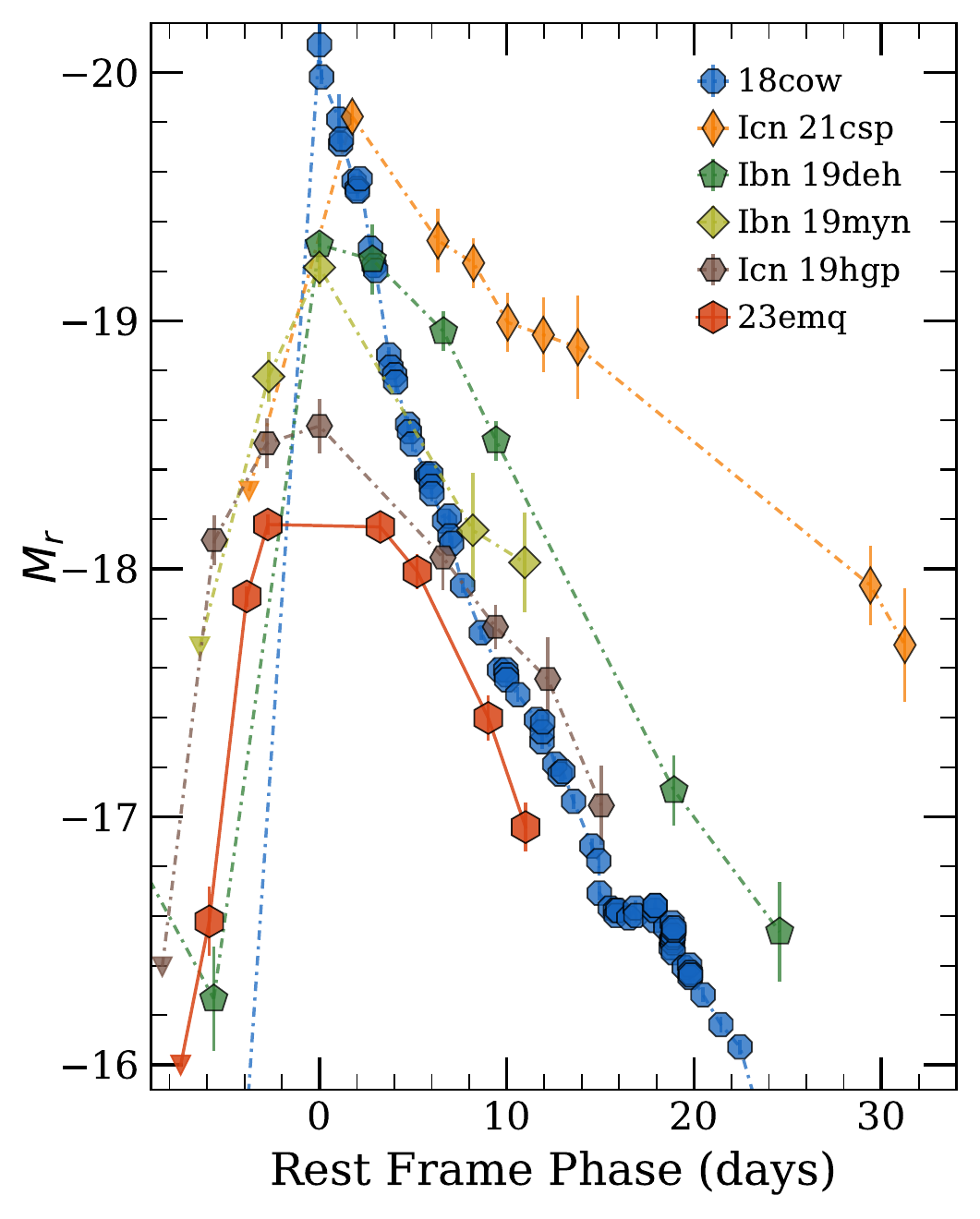}
        \end{minipage}
        \begin{minipage}[b]{0.325\textwidth}
            \centering
            \includegraphics[width=\textwidth]{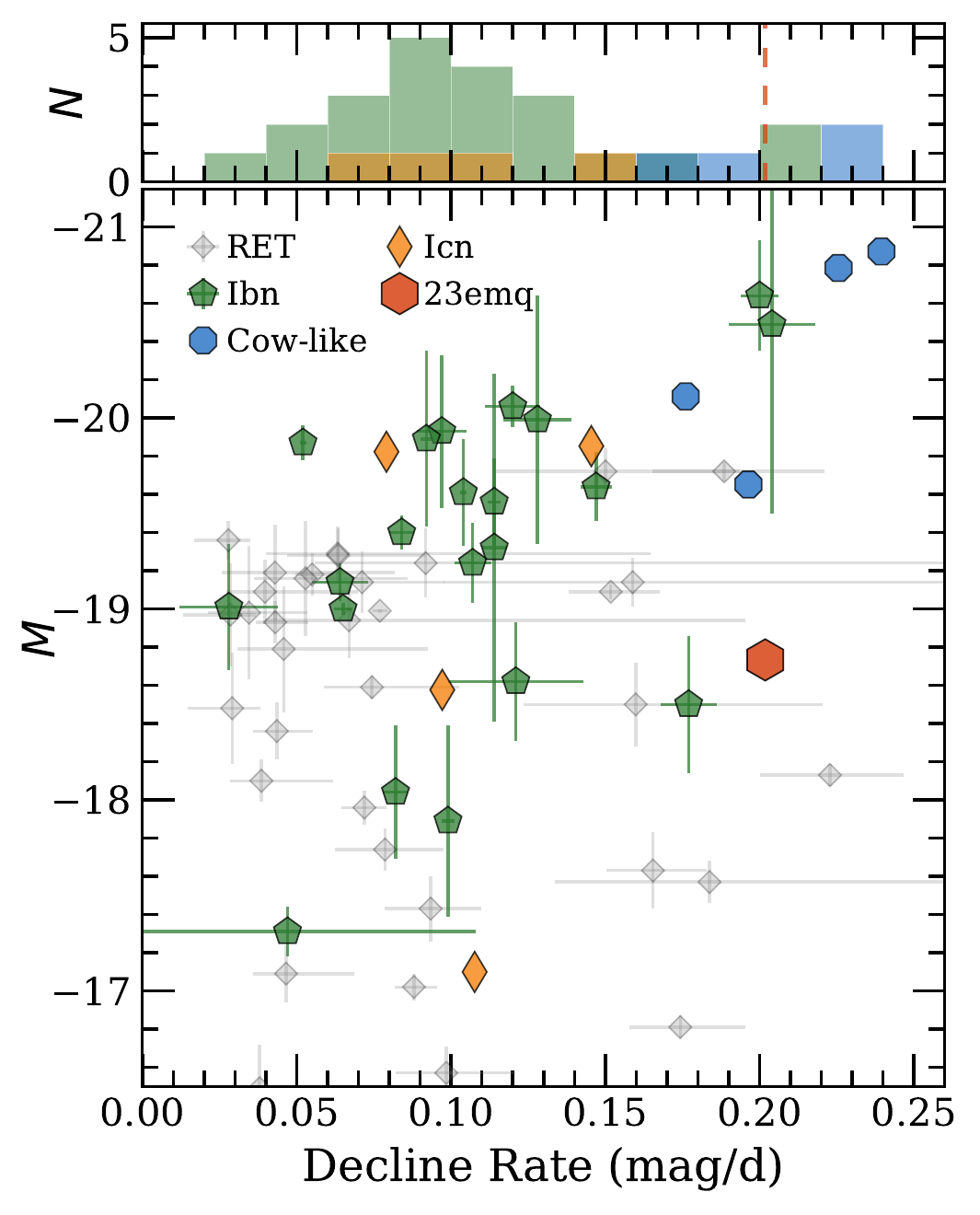}
        \end{minipage}
        \begin{minipage}[b]{0.325\textwidth}
            \centering
            \includegraphics[width=\textwidth]{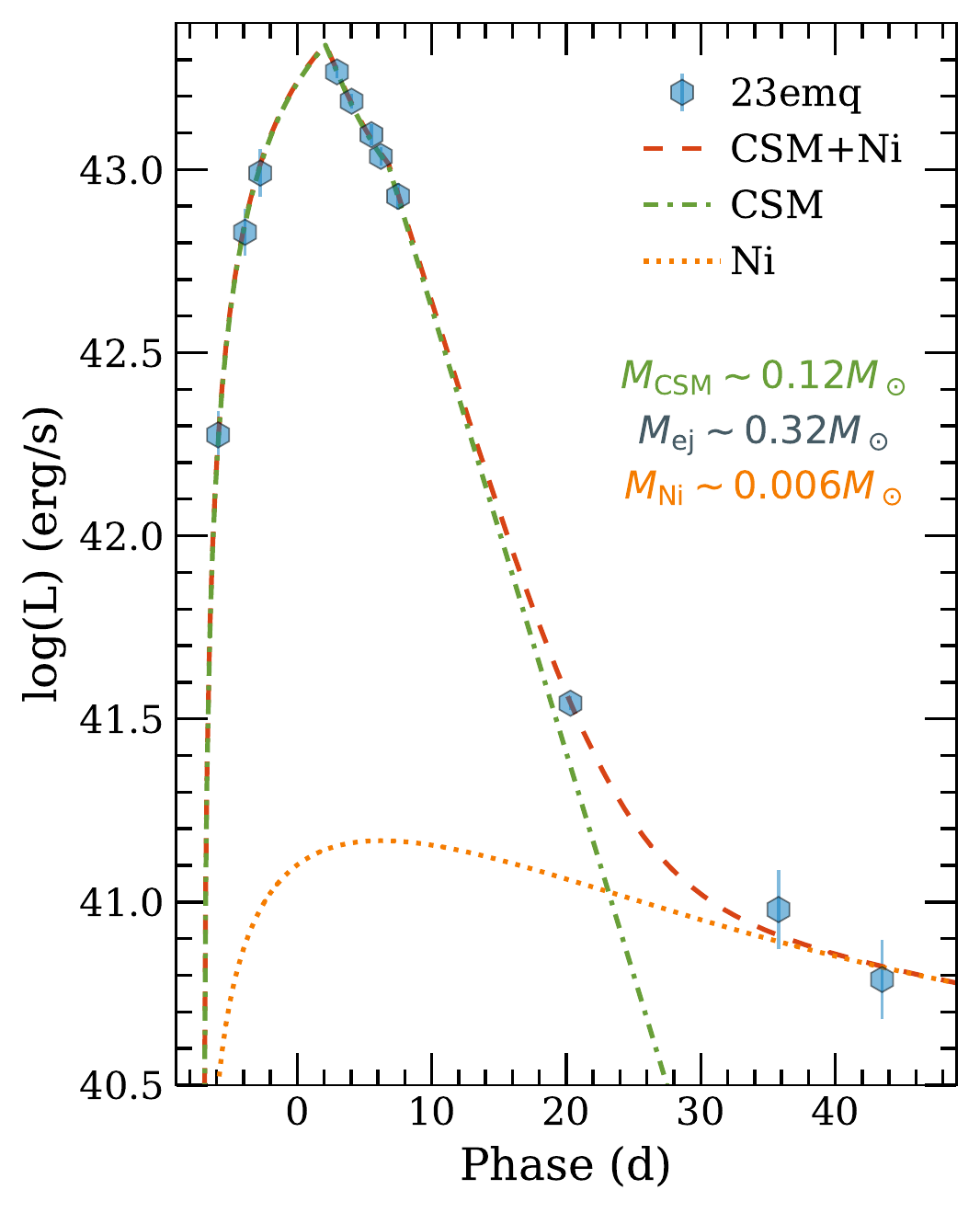}
        \end{minipage} %
   \caption{Photometric qualities of SN\,2023emq. Top: The multi-band light curve of SN\,2023emq. Open markers refer non-detections ($5\sigma$). The spectral epochs are highlighted with dashed lines. The interpolated $o$-band GP light curve is shown over the data, and the decline rate of $^{56}$Co is drawn for comparison.The late-time decline rate is significantly faster than the  $^{56}$Co rate, and the decline continued fast until the SN was no longer detected at $+78$\,d and $+95$\,d. Left: The ATLAS $o$-band light curve of SN\,2023emq in comparison to $r$-band light curves of example fast events. The light curves are collected from ZTF via the Alerce broker \citep{Forster2021} except for AT\,2018cow where data presented in \citet{Prentice2018,Perley2018} was used.  Middle: Absolute magnitude vs. decline rate of fast-evolving transients. Values of Type Icn SNe \citep{Pellegrino2022}, cow-like transients \citep{Prentice2018,Perley2018,Ho2020,Perley2021, Yao2022} and SN\,2023emq were obtained by fitting the first 20\,d  of post-peak $r$-band light curves linearly. The values for RETs are taken from \citet{Pursiainen2018}, and the used bands ($g$, $r$, $i$ or $z$) were chosen to be the closest to the rest frame $r$-band. Ibn SNe are collected from \citet{Hosseinzadeh2017}, where data are presented mostly in $R$-band. On the top histogram we show the number (N) of transients per decline rate with SN\,2023emq highlighted with dashed red line. No Icn SNe and only 2/22 Ibn are faster than SN\,2023emq. Right: The best-fitting combined CSM and nickel decay model \citep{Chatzopoulos2012, Chatzopoulos2013} to the bolometric light curve of SN\,2023emq. The derived CSM and ejecta masses are similar, but smaller in comparison to other Ibn/Icn SNe as expected given the fast light curve evolution near peak. The nickel mass is comparable to fast-evolving Ibn/Icn SNe.} 
    
    \label{fig:LC_prop}
\end{figure*}

\section{Observations and data reduction}
\label{sec:obs}

SN\,2023emq was discovered by Asteroid Terrestrial-impact Last Alert System \citep[ATLAS;][]{Tonry2018,Smith2020} on 2023 Apr 01 under the name ATLAS23ftq, with a previous non-detection two days prior \citep{Tonry2023}. The SN is associated with LEDA797708 \citep{Makarov2014}, a visibly blue galaxy in Pan-STARRS1 $3\pi$ images, with an offset of $\sim3.7\arcsec$ from the brightest core region \citep{Flewelling2020}. Late-time image of the environment obtained with FOcal Reducer and low dispersion Spectrograph \citep[FORS2;][]{Appenzeller1998} on the Very Large Telescope (VLT) at European Southern Observatory (ESO), Paranal, Chile, is shown in Figure \ref{fig:VLT_stamp}.

The photometry of SN\,2023emq was collected from the public ATLAS and Zwicky Transient Facility \citep[ZTF;][]{Bellm2019a} surveys, and with Target-of-opportunity observations with the Neil Gehrels Swift Observatory UV-Optical telescope (UVOT, PIs: Brown and Pellegrino). The late-time ($\gtrsim+20$\,d) evolution was monitored with aperture photometry from Alhambra Faint Object Spectrograph and Camera (ALFOSC) mounted on the Nordic Optical Telescope at La Palma, Spain (PI: Pursiainen) and from FORS2/VLT with a Director's Discretionary Time (DDT) Programme (2111.D-5006; PI: Pursiainen). The ATLAS $o$ and $c$ band light curves were generated using the ATLAS Forced Photometry service\footnote{\url{fallingstar-data.com/forcedphot/}} \citep{Shingles2021}. The ZTF $g$ and $r$ band light curves were collected using the Alerce \citep{Forster2021} and Lasair \citep{Smith2019} brokers. The light curves from Swift UVOT were reduced following \citet{Charalampopoulos2022a}. The photometric data are provided in Table \ref{tab:all_phot}.

The spectra of SN\,2023emq were obtained with NOT/ALFOSC, ESO Faint Object Spectrograph and Camera 2 (EFOSC2) on the New Technology Telescope (NTT) at La Silla observatory, Chile by the extended Public ESO Spectroscopic Survey for Transient Objects plus \citep[ePESSTO+;][]{Smartt2015} survey and VLT/X-Shooter \citep{Vernet2011} under the DDT programme. The NOT spectrum was reduced using the \texttt{PyNOT-redux} reduction pipeline,\footnote{\url{github.com/jkrogager/PyNOT/}} the NTT spectrum with the PESSTO pipeline \citep{Smartt2015} and the X-Shooter spectrum as described in \citet{Selsing2019}. We also analysed the Global SN Project classification spectrum \citep{Pellegrino2023} available in WISeREP\footnote{\url{www.wiserep.org}} \citep{Yaron2012}. The spectral log is provided in Table \ref{tab:spec_log}. Finally, we obtained one epoch of NOT/ALFOSC $V$-band imaging polarimetry on 2023 Apr 14 ($+8.7$\,d). The data was reduced following \citet{Pursiainen2023a}.

\begin{figure*}
    \centering
    \includegraphics[width=0.98\textwidth]{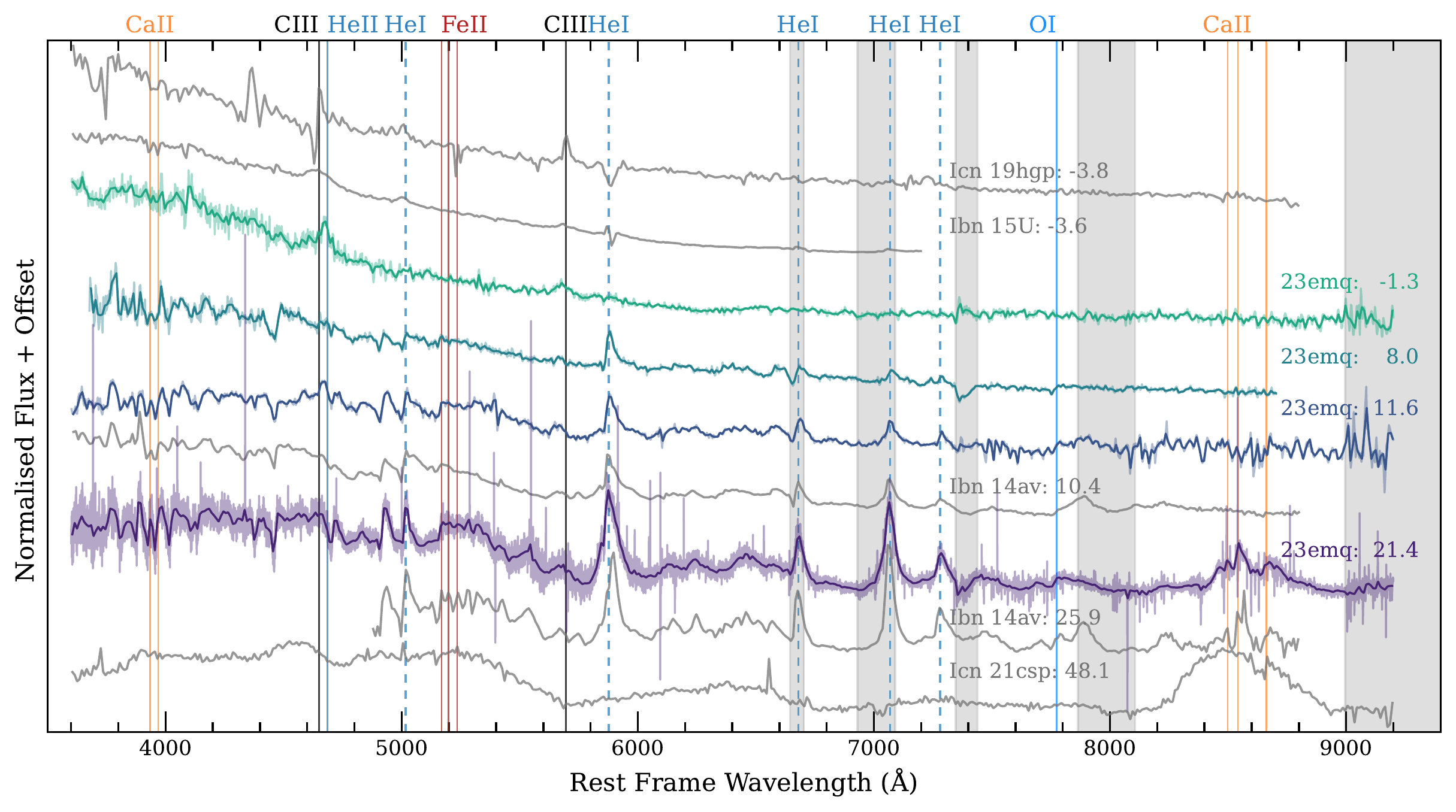}
    \caption{The spectroscopic time series of SN\,2023emq (in colours) in comparison to the Type Icn SNe 2019hgp \citep{Gal-Yam2022} and 2021csp \citep{Perley2022}, and the Type Ibn SNe 2015U \citep{Shivvers2016} and 2014av \citep{Pastorello2016} in grey. Spectral epochs are presented with respect to peak brightness. The first spectrum resembles that of SN\,2015U and the latest spectrum is virtually identical to SN\,2014av. As already noted by \citet{Perley2022}, the continuum emission of SN\,2021csp at late times is very similar to the continuum seen in Ibn SNe including SN\,2023emq. The most prominent \ion{He}{1} features are marked with dashed lines, other strong features with solid lines, and tellurics are shown in grey bands. Note that the spectrum of the highly reddened SN\,2015U is artificially de-reddened for visual comparison.}
    \label{fig:spectral_timeseries}
\end{figure*}

\section{Analysis}
\label{sec:analysis}

\subsection{Photometry}
\label{subsec:photometry}

The multi-band light curve of SN\,2023emq is shown in Figure \ref{fig:LC_prop} (top). While the decline is well covered, only the ATLAS $o$-band has pre-peak data. To estimate the peak MJD and magnitude we use Gaussian process (GP) interpolation using the setup presented in \citet{Pursiainen2020} and find the peak to be at $\mathrm{MJD}\sim60040$ and $-18.7\pm0.1$\,mag. To highlight the extreme evolution timescale of SN\,2023emq, we compare its light curve to the enigmatic AT\,2018cow and to example Icn/Ibn SNe chosen for their fast light curve evolution in Figure \ref{fig:LC_prop} (left). While the rise of the SN is not exceptional, its initial decline is fast regardless of its type, and by the time of the last spectrum at $\sim+20$\,d, the SN had already declined by 3.5\,mag in brightness. In fact, the decline rate in the first 15 days post-peak ($\sim0.20$\,mag/d) is at the extreme end of Ibn/Icn SNe and comparable to the events similar to AT\,2018cow \citep[e.g.][]{Prentice2018,Perley2018,Ho2020,Perley2021, Yao2022}, as demonstrated in Figure \ref{fig:LC_prop} (middle). However, based on our NOT $R$-band data we can determine that after the observation at $+20$\,d the decline slowed to $0.07$\,mag/d --  three times slower than the early decline. While the decline rates of Ibn SNe are expected to decrease in time and similar breaks have been seen before \citep[e.g. SN\,2015G;][]{Shivvers2017a}, such extreme transitions are rare in the population \citep[see e.g.][]{Hosseinzadeh2017}. Out of the Type Icn SNe, only SN\,2022ann has exhibited a similar transition where the decline rate decreased from a maximum of $\sim0.14$\,mag/d  to $\sim0.03$\,mag/d \citep{Davis2022}. 

The late-time decline of SN\,2023emq is faster than that expected from the $^{56}$Co to $^{56}$Fe decay (0.0098\,mag/d), but the decay likely contributes at these epochs. Using the semi-analytical light curve models of \citet{Chatzopoulos2012,Chatzopoulos2013}, we fit the bolometric light curve of SN\,2023emq (for details see Appendix \ref{sec:L_bol}) with a combined CSM-interaction and nickel decay model.  We assumed shell-like CSM with a setup presented by \citet{Pellegrino2022} for type Icn SNe except for the opacity for which we used $\kappa=0.2$\,cm$^{2}$s$^{-1}$ as appropriate for H-poor CSM \citep{Chatzopoulos2013}. As shown in Figure \ref{fig:LC_prop} (right), we find a good fit with $M_\mathrm{CSM}=0.12_{-0.01}^{+0.01}M_\odot$ and $M_\mathrm{ejecta}=0.32_{-0.04}^{+0.04}M_\odot$ ($1\sigma$) with  $M_\mathrm{Ni}=0.006_{-0.001}^{+0.001}M_\odot$. The recovered values of $M_\mathrm{CSM}$ and $M_\mathrm{ejecta}$ are smaller than found in the literature for Type Ibn SNe \citep[e.g.][]{Pellegrino2022a}, as expected given the fast photometric evolution. The $M_\mathrm{Ni}$ is similar to the lowest estimates for Type Ibn/Icn SNe \citep[e.g.][]{Pellegrino2022}. However, based on our late-time non-detections the decline continued furiously. The SN was no longer detected in host-subtracted NOT $R$ band photometry taken at $+78$\,d with a $5\sigma$ upper limit of $22.7$\,mag. A deeper limit of $25.1$\,mag was obtained at $+95$\,d with VLT. As the decline continued faster than expected of $^{56}$Co decay, the amount of $^{56}$Ni synthesised in the explosion has to be truly marginal.

\subsection{Spectroscopy}
\label{subsec:spectroscopy}

The spectral time series of SN\,2023emq is compared to example Type Ibn and Icn SNe, chosen based on their spectral similarity to SN\,2023emq, in Figure \ref{fig:spectral_timeseries}. We adopt a redshift of $z=0.0338$, determined based on host galaxy emission lines present in the 2D X-Shooter spectrum. In the classification spectrum, the most notable emission lines at $5700$\,Å and $4690$\,Å are identified as \ion{C}{3} $\lambda5696$ and a blend of \ion{C}{3} $\lambda4650$ and \ion{He}{2} $\lambda4686$ \citep{Pellegrino2023}. These lines are common in Type Icn SNe, as demonstrated by well-studied SN\,2019hgp \citep{Gal-Yam2022}, but emission at similar wavelengths has also been seen in a few Type Ibn SNe (e.g. SN\,2015U). In the subsequent spectra the SN has evolved into a typical Type Ibn SN characterised by strong \ion{He}{1} emission lines. The spectra are virtually identical to that of Ibn SN\,2014av shown in the figure. 

The spectral evolution of the SN suggests that it could be a transitional Type Icn/Ibn SN, but the question relies on whether the early spectroscopic signatures are also seen in Type Ibn SNe, and important SNe to compare SN\,2023emq to are the flash-ionised Type Ibn SNe 2010al \citep{Pastorello2015c}, 2019uo \citep{Gangopadhyay2019} and 2019wep \citep{Gangopadhyay2022}. SNe often exhibit short-lived flash-ionised features early in their evolution as a result of ionisation of the nearby CSM by the supernova shock breakout \citep[e.g.][]{Gal-Yam2014a}. All three SNe exhibit a strong emission feature at $4650$\,Å due to a blend of \ion{N}{3}, \ion{C}{3} and \ion{He}{2}, as is typical for flash-ionised spectra of young SNe \citep[see e.g. ][]{Khazov2016, Bruch2021, Bruch2023}. However, while no emission is present at $5700$\,Å in SN\,2010al, in SN\,2019uo and SN\,2019wep extremely faint \ion{C}{3} $\lambda5696$ emission is reported. This is also the case for H-rich flash-ionised spectra, as the line is typically extremely faint if seen at all \citep[e.g. SN\,1998S;][]{Fassia2001}. Furthermore, \citet{Shivvers2016} showed that a similar but stronger emission line at $5700$\,Å was also seen in an early spectrum of Type Ibn SN\,2015U in addition to the blended emission line at $4650$\,Å (see Figure \ref{fig:spectral_timeseries}). The authors suggested that the line is \ion{N}{2} $\lambda5680$, supported by the presence of several \ion{N}{2} absorption lines in the later spectra \citep{Pastorello2015a, Shivvers2016}. Regardless of the exact identification of the line in SN\,2023emq, emission has been detected at the two wavelengths in both Ibn and Icn SNe, and the presence of emission alone does not warrant a Icn classification.

\begin{figure*}
    \centering
        \begin{minipage}[b]{0.49\textwidth}
            \centering
            \includegraphics[width=\textwidth]{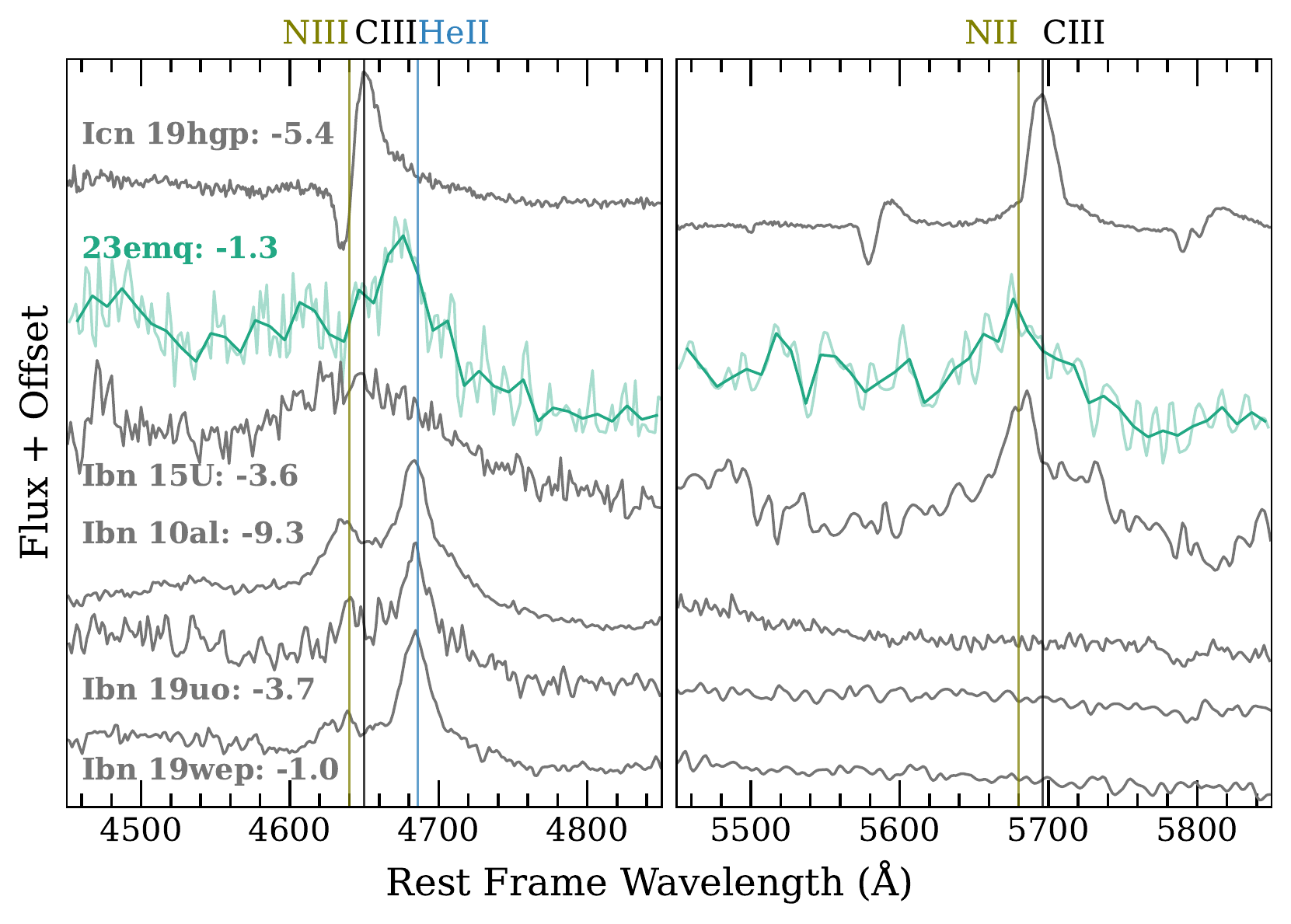}
        \end{minipage} %
        \begin{minipage}[b]{0.49\textwidth}
            \centering
            \includegraphics[width=\textwidth]{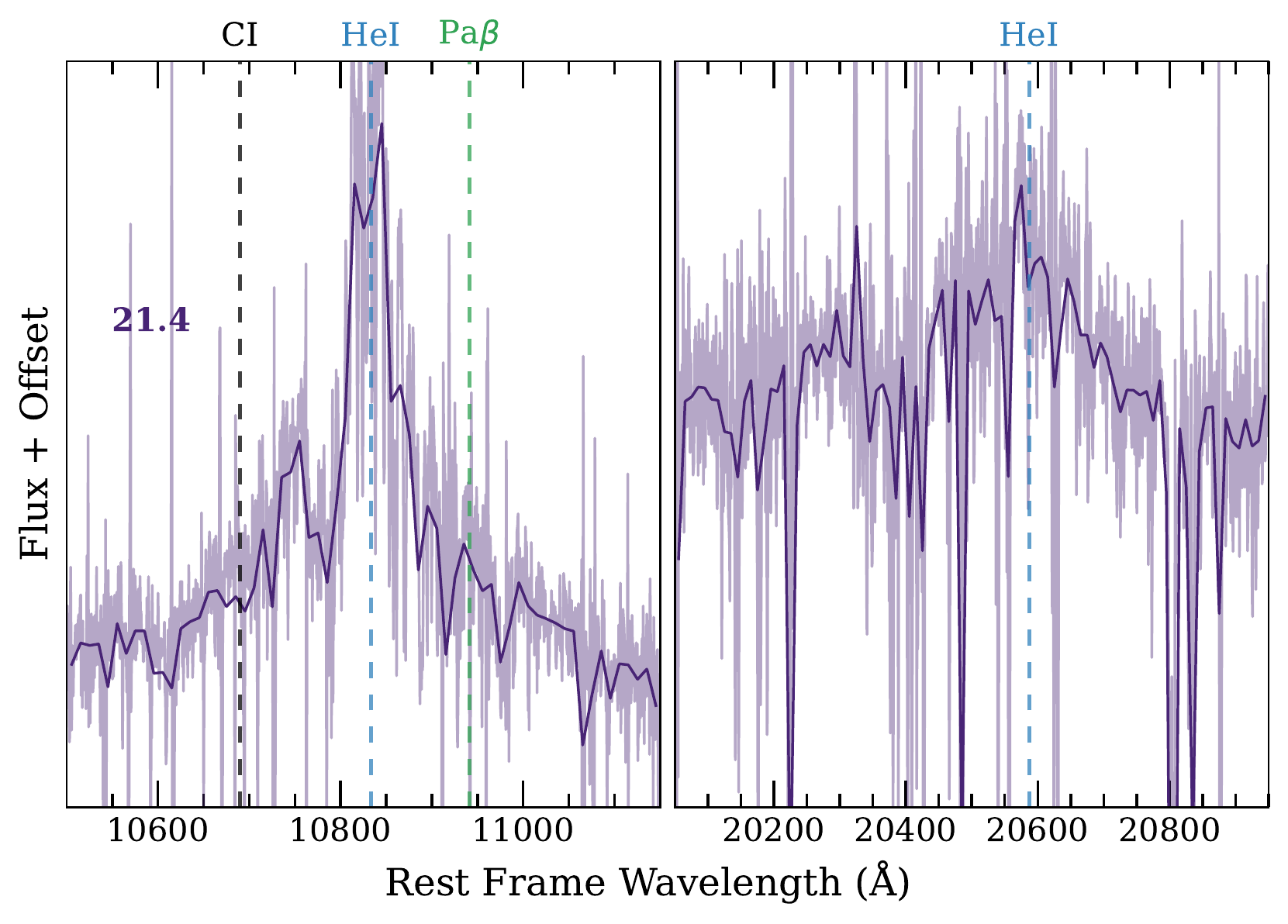}
        \end{minipage}
    \caption{Left: The classification spectrum of SN\,2023emq with the possible line identifications. The feature at $4690$\,Å is dominated by \ion{He}{2} $\lambda4686$ with faint blue excess possibly caused by either \ion{N}{3} $\lambda4640$ or \ion{C}{3} $\lambda4650$. At $5700$\,Å the emission could be either \ion{N}{2} $\lambda5680$ or \ion{C}{3} $\lambda5696$. The spectrum of SN\,2015U from Figure \ref{fig:spectral_timeseries} as well as early spectra of the Type Icn SN\,2019hgp \citep{Gal-Yam2022} and the  flash-ionised Ibn SNe 2010al, 2019uo, and 2019wep \citep{Pastorello2015c, Gangopadhyay2019, Gangopadhyay2022} are shown for comparison. Emission is seen in all SNe around $4650$\,Å. At $5700$\,Å the emission feature is either faint (SN\,2015U) or not clearly detected (SNe 2010al, 2019uo, 2019wep) in the Ibn SNe, but it is prominent in the Type Icn SN\,2019hgp.  Right: The only notable features in the X-Shooter NIR spectrum. \ion{He}{1} $\lambda10830$ is clear, but the strong \ion{C}{1} emission line $\lambda10690$ seen in Type Icn SN\,2021csp \citep{Fraser2021} is not present. \ion{He}{1} $\lambda20587$ is also tentatively identified.}
    \label{fig:spec_lines}
\end{figure*}

\begin{figure}
    \centering
    \includegraphics[width=0.48\textwidth]{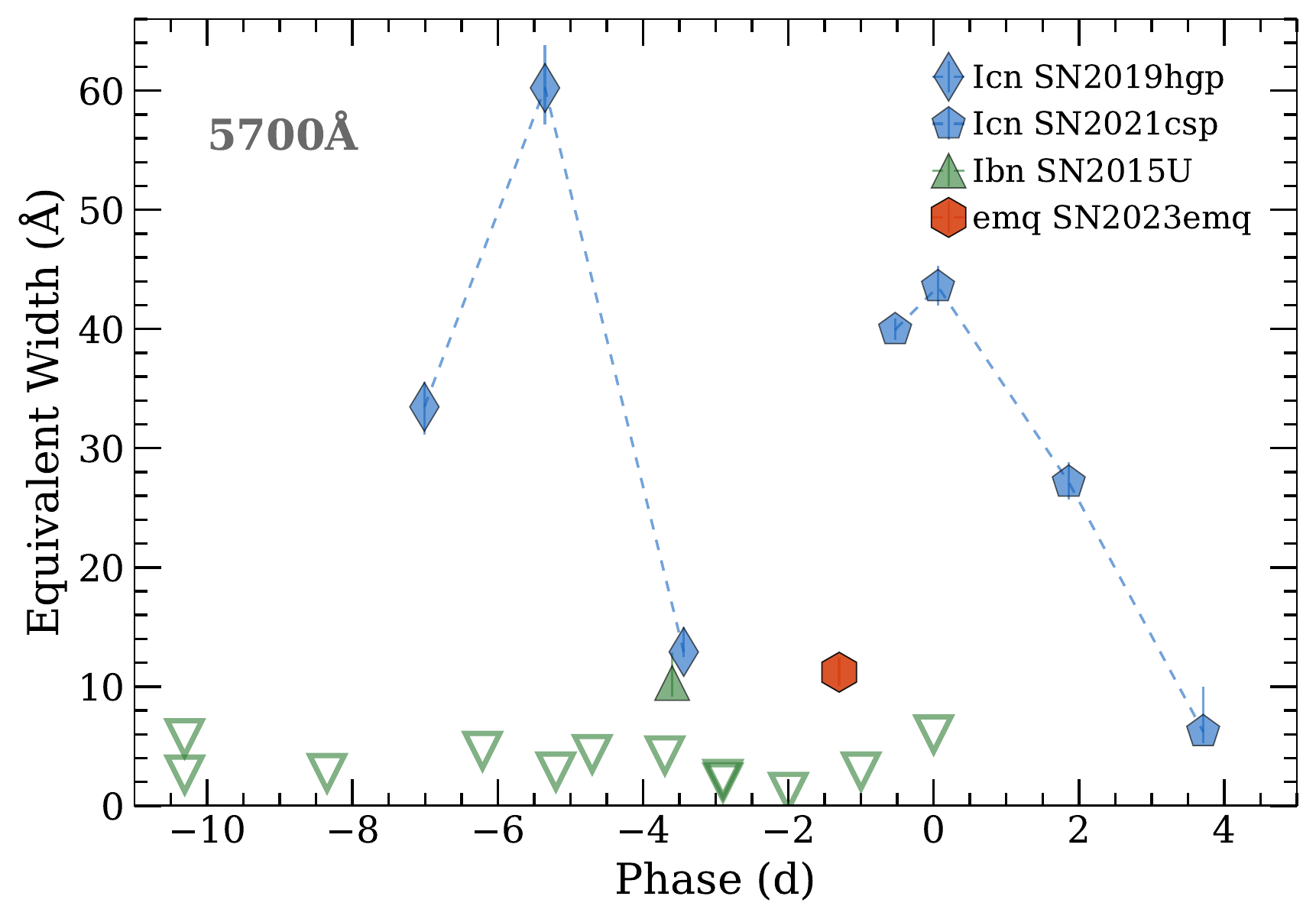}
    \caption{Equivalent width (EW) of the emission at $5700$\,Å as function of rest frame phase for SN\,2023emq, Icn SNe 2019hgp and 2021csp and flash-ionised Type Ibn SNe. For Ibn SNe, the emission feature is clearly seen in only SN\,2015U (see Figure \ref{fig:spec_lines}), and for the other Ibn spectra we show $3\sigma$ upper limits (open triangles) determined assuming a Gaussian profile with FWHM measured for SN\,2023emq ($\sim3000$\,km/s). SN\,2023emq does not exhibit as prominent emission as seen in most Icn spectra, but the line is also far more significant than seen in most flash-ionised Ibn SNe, possibly implying a continuum of properties between the two classes. The EWs were estimated with Gaussian fits over a linear background. For Ibn SNe we used the spectra of SN\,2010al, SN\,2015U, SN\,2019uo and SN\,2019wep that exhibited the strong blended emission at $4650$\,Å -- indicative of the flash-ionised phase. For the Icn SN\,2019hgp and SN\,2021csp we used the spectra that showed \ion{C}{3} $\lambda5696$ emission, excluding the lower resolution LT/SPRAT spectra  \citep[for details see][]{Gal-Yam2022, Perley2022}. }
    \label{fig:ews}
\end{figure}

In Figure \ref{fig:spec_lines} (left), we investigate what emission lines are present in the first spectrum in comparison to Ibn and Icn SNe. The line that best matches the feature at $4690$\,Å is \ion{He}{2}, with a fainter contribution from \ion{N}{3}/\ion{C}{3} on the blue side of the feature. The presence of \ion{He}{2} is indicative of Type Ibn SN, as in Type Icn the line is not prominent \citep[e.g.][]{Pellegrino2022}. Furthermore, the feature of Ibn SN\,2015U appears to have the same shape as in SN\,2023emq if the \ion{He}{2} contribution is ignored while in Icn SNe \ion{C}{3} $\lambda4650$ typically shows a P~Cygni profile not seen in SN\,2023emq. At $5700$\,Å the line is well-matched with both \ion{C}{3} $\lambda5696$ and \ion{N}{2} at $5680$\,Å and cannot be clearly distinguished.  In comparison to Ibn SNe, only SN\,2015U exhibits any clear emission in the interval, and in this regard SN\,2023emq appears more similar to Icn SNe, where the emission is prominent. The trend is also visible in Figure \ref{fig:ews} where we show the equivalent widths (EWs) of emission at $5700$\,Å for early spectra of Type Icn SNe 2019hgp and 2021csp and the flash-ionised Ibn SNe against SN\,2023emq. The EW of SN\,2023emq ($11.2_{-1.1}^{+1.2}$\,Å) is smaller than in most of the Icn spectra, but comparable to the ones measured at later phases just before \ion{C}{3} emission faded away \citep[e.g.][]{Perley2022}. On the other hand, SN\,2023emq is similar to SN\,2015U, but clearly above the $3\sigma$ upper limits (open triangles) estimated for the other Ibn SNe.

Finally, we searched the later spectra of SN\,2023emq for signatures of \ion{C}{2} or \ion{C}{1} as seen in Icn SNe,  but found no clear features. In the X-Shooter NIR spectrum, taken at $+21.4$\,d, we identify a clear \ion{He}{1} $\lambda10830$ emission line (\ref{fig:spec_lines}, right), but do not see the strong \ion{C}{1} $\lambda10690$, detected in Icn SN\,2021csp at a similar epoch \citep[$+20.5$\,d;][]{Fraser2021}. 

Given the presence of the prominent \ion{He}{2} emission in the first spectrum, the early spectral similarity to SN\,2015U and the flash-ionised Type Ibn SNe as well as the unambiguous Ibn nature after peak brightness, we favour the interpretation that SN\,2023emq is a flash-ionised Type Ibn SN. The unusually prominent emission line at $5700$\,Å might imply some kind of continuum between Icn and Ibn SNe and could indicate that the SN\,2023emq is between the two populations if it is related to \ion{C}{3}. However, due to the unclear nature of the line, it is possible that it is related to \ion{N}{2} like suggested for SN\,2015U \citep{Shivvers2016}. In this case, SN\,2023emq is the second example of Ibn SN with \ion{N}{2} emission with no obvious connection to Type Icn SNe.

We also used the X-Shooter spectrum to investigate the effect of host galaxy extinction. The Na\,ID doublet, which is typically used to estimate extinction in a galaxy, is not clearly visible in our high-resolution X-Shooter spectrum, and we can estimate the upper limit for $E_\mathrm{B-V}$ of the host by generating Gaussian absorption profiles over the spectrum. Assuming that the full width at half maximum of the line is the resolution of the spectrograph ($35$\,km/s at 5890\,Å), we find a $3\sigma$ limit for the total equivalent width of the doublet is $\mathrm{EW}\sim0.146$\,Å. Using the empirical equation of \citet{Poznanski2012}, this equates to $E_\mathrm{B-V}\lesssim0.02$, indicating that the effect of the host galaxy extinction should be negligible. 

\subsection{Polarimetry}
\label{subsec:polarimetry}

\begin{figure*}
    \centering
    \begin{minipage}[b]{0.48\textwidth}
        \centering
        \includegraphics[width=\textwidth]{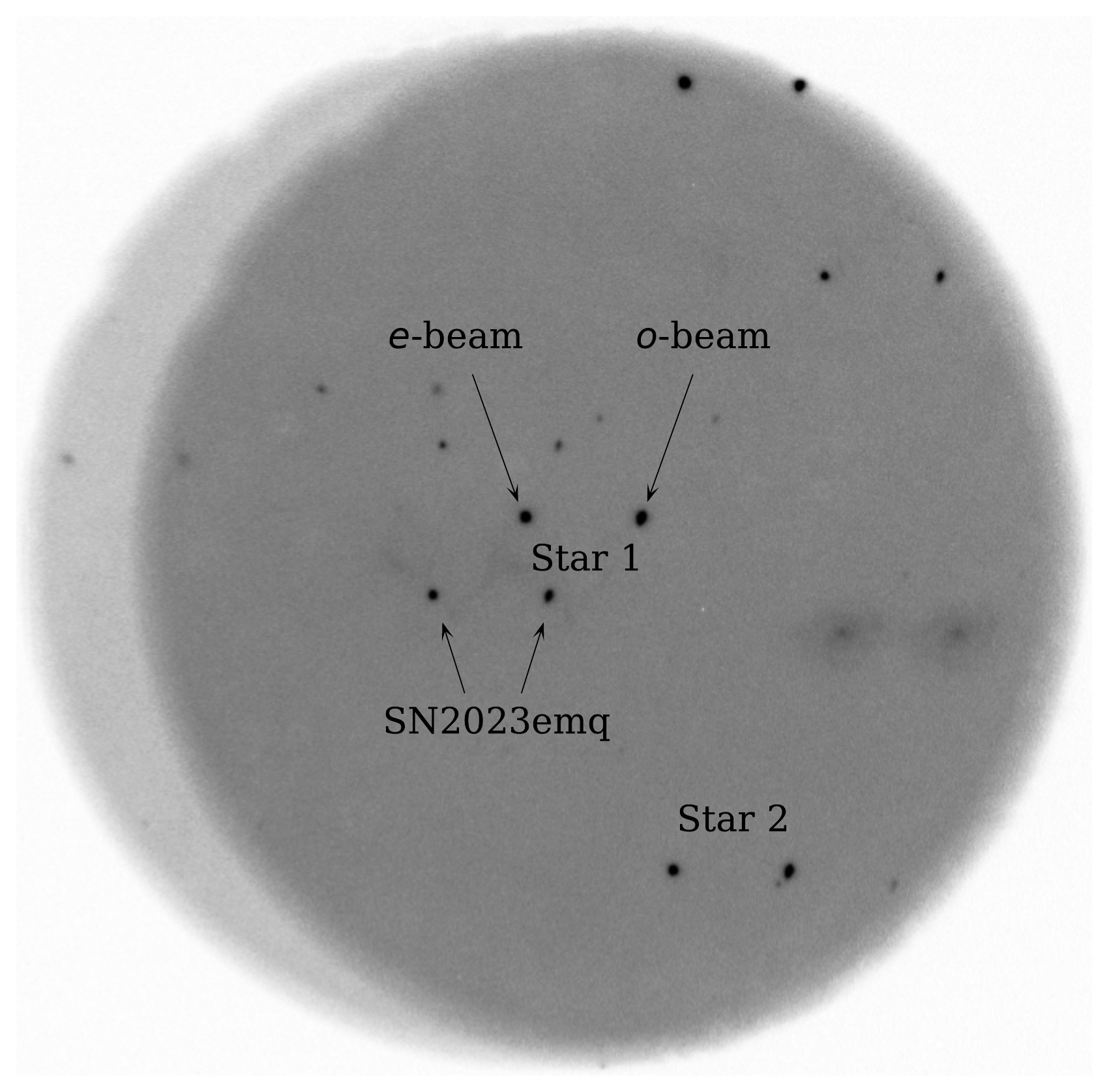}
    \end{minipage} %
    \begin{minipage}[b]{0.48\textwidth}
        \centering
        \includegraphics[width=\textwidth]{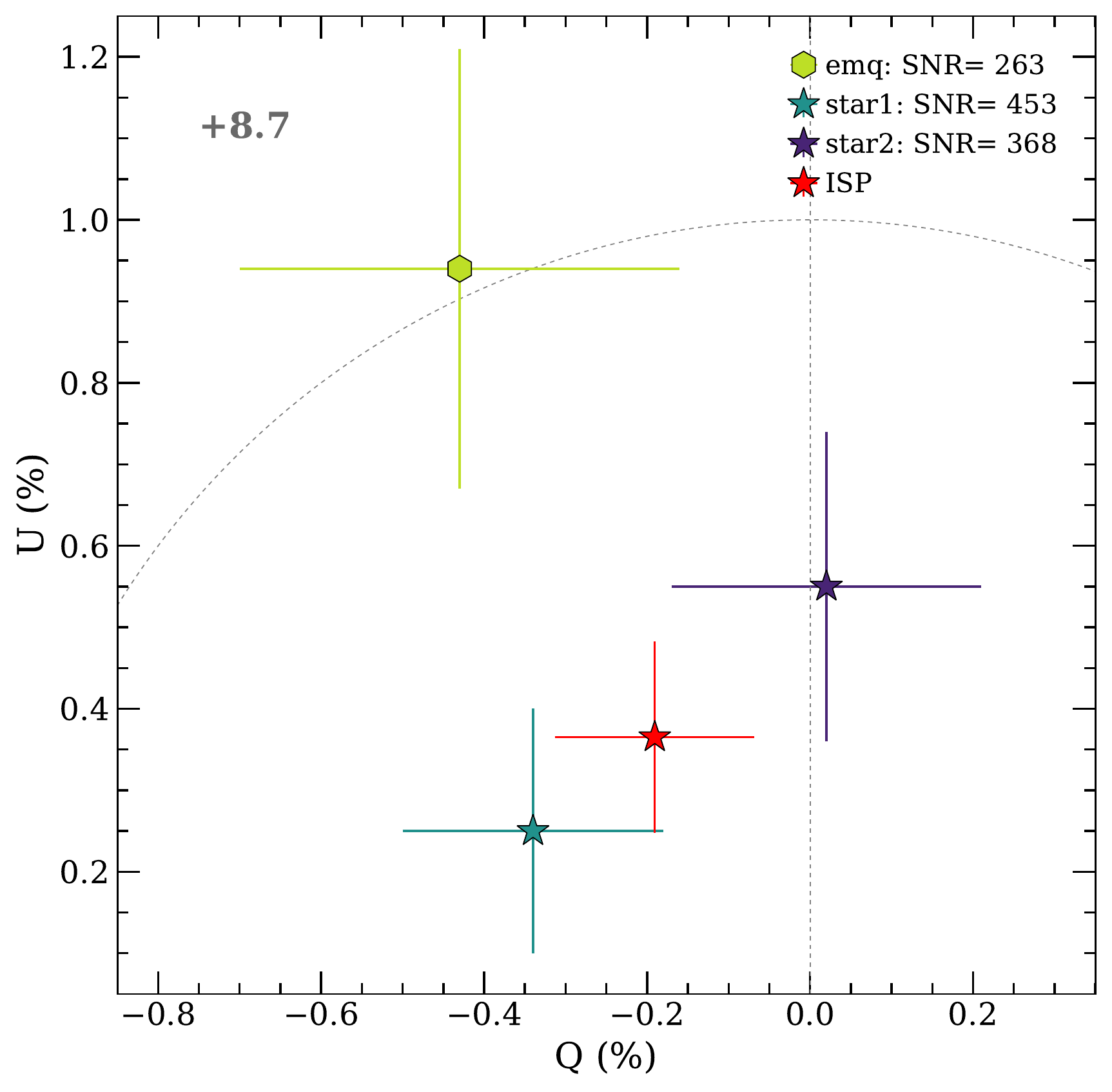}
    \end{minipage} %
    \caption{The NOT $V$-band polarimetry of SN\,2023emq taken at $+8.7$\,d post-peak. Left: The field of view of the observation. In the polarimetric mode, the extraordinary ($e$) and ordinary ($o$) beams are overlaid 15$\arcsec$ apart. The SN and the two bright stars are marked. Other sources in the image were either too faint or too close to the edge for reliable estimation of the Stokes parameters. Right: The Stokes $Q$\,--\,$U$ plane. Both the SN and the two stars found in the image are shown. The SN is clearly offset from the stars. After correcting for the Galactic ISP and polarisation bias, we find intrinsic polarisation of $P=0.55\pm0.30$\%. The dashed lines mark the location of $Q=0$\% and $P=1$\%.}
    \label{fig:NOT_pola}
\end{figure*}

The polarimetry taken at $+8.7$\,d shows that SN\,2023emq is polarised as demonstrated in Figure \ref{fig:NOT_pola} (right). We measure the dimensionless Stokes parameters to be $Q=-0.43\pm0.27$\% and $U=0.94\pm0.27$\% corresponding to $P=1.03\pm0.27$\%. In addition, it is possible to obtain measurements for two stars found in the ALFOSC field of view shown in Figure \ref{fig:NOT_pola} (left). Both stars are bright, isolated targets, and based on the astrometric solutions from Gaia Data Release 3 \citep{GaiaCollaboration2022} they also lie $150$\,pc above the Milky Way plane and should therefore be sufficiently far to probe the Milky Way dust content \citep{Tran1995}. As such, they should be reliable estimators of the Galactic interstellar polarisation (ISP). The SN is clearly offset from the stars on the $Q$\,--\,$U$ plane (Figure \ref{fig:NOT_pola}), which is an indication of intrinsic polarisation. Assuming the Galactic ISP can be obtained by averaging the two stars, we find Galactic ISP-corrected values for SN\,2023emq of $Q=-0.24\pm0.30$\% and $U=0.37\pm0.30$\%, resulting in $P=0.55\pm0.30$\% \citep[corrected for polarisation bias following ][]{Plaszczynski2014}. 

While we have no means to directly probe the host galaxy ISP, its maximum value should correlate with the host extinction following $P_{\mathrm{ISP}} < 9 \times E_\mathrm{B-V}$ \citep{Serkowski1975}. Using the derived limit on host extinction ($E_\mathrm{B-V}\lesssim0.02$) we find that the effect of host galaxy dust content should be $P_{\mathrm{ISP, Host}}\lesssim0.19$\%. As the derived polarisation degree ($P\sim0.55$\%) is more significant, it is likely intrinsic to the SN.  Given the spectrum at the time does not exhibit strong absorption lines that could induce polarisation \citep[e.g.][]{Wang2008}, the polarisation is likely that of the continuum. If the photosphere is an oblate ellipsoid, $P\sim0.55$\% corresponds to a lower limit of the physical minor over major axial ratio is $b/a\lesssim0.9$ \citep{Hoflich1991}, but as shown by \citet{Pursiainen2022} the CSM can also be found in a disk/torus, and the CSM does not need to be in a uniform ellipsoidal shape. However, given the detection is not very strong, we conclude that the CSM shows high degree of spherical symmetry, in perhaps marginally aspherical configuration. 

Polarimetry of Type Ibn/Icn SNe is very scarce in the literature, and polarimetric data that probes the intrinsic properties of the SNe has been presented only for one Ibn and one Icn SN. The spectral polarimetry obtained for Type Ibn SN\,2015G $5$\,d after discovery showed $P\sim2.7$\%, indicative of high asymmetry \citep{Shivvers2017a}, while the one epoch for Type Icn SN\,2021csp at $+3.5$\,d shows low polarisation, implying high spherical symmetry \citep{Perley2022}. Additionally, \citet{Shivvers2016} presented three epochs of spectropolarimetry of the highly reddened Type Ibn SN\,2015U at $\sim+5$\,d and conclude that the observed polarisation signal was dominated by the contribution of dust in the host galaxy, and no firm conclusions on the polarisation of the SN were made. As such, our broad-band polarimetry for SN\,2023emq is only the third observation that constrains the photospheric geometry of Ibn/Icn SNe. The H-poor CSM appears to often show high degree of spherical symmetry,  but more events need to be observed to investigate the distribution of their photospheric shape.

\section{Conclusions}
\label{sec:conclusions}

We have presented an analysis of the photometric, spectroscopic and polarimetric properties of the fast evolving H-poor interacting SN\,2023emq. While the rise of the SN is not particularly fast in the context of Type Ibn and Icn SNe, its initial decline ($\sim0.20$\,mag/d) is remarkable and even comparable to AT\,2018cow. After $+20$\,d the decline rate slowed significantly to ($\sim0.07$\,mag/d). Such a large transition in the decline rate is extreme and to our knowledge has not been seen before in interacting H-poor SNe.  We modelled the bolometric light curve with a combined CSM interaction and nickel decay model and find good matches with $M_\mathrm{ejecta}\sim0.32M_\odot$ and $M_\mathrm{CSM}\sim0.12M_\odot$, with $\sim0.006M_\odot$ of $^{56}$Ni. While the late-time light curve is still faster than expected of $^{56}$Co to $^{56}$Fe decay, it is likely that the radioactive decay is playing a part.

The SN was initially classified as Type Icn based on the prominent emission features at $4650$\,Å and $5700$\,Å identified as \ion{C}{3} as typical in Icn SNe \citep[e.g.][]{Gal-Yam2022, Perley2022}, but by $+8$\,d the SN had evolved into a typical Type Ibn characterised by prominent, narrow \ion{He}{1} emission features, possibly implying a transitional nature. We, however, conclude that SN\,2023emq is flash-ionised Ibn SN instead. 
As emission at $4650$\,Å and $5700$\,Å has been seen in a few Type Ibn SNe, just the presence of emission is not yet enough to classify the SN as Type Icn. The emission at $4650$\,Å seen in the first spectrum of SN\,2023emq is dominated by \ion{He}{2} with contribution from \ion{C}{3}/\ion{N}{3}, indicative of Ibn nature. Additionally, the \ion{C}{3} $\lambda4650$ has typically a prominent P~Cygni profile in Type Icn SNe, but only emission is present in SN\,2023emq. However, while emission at $5700$\,Å has been reported in some Type Ibn SNe, it is remarkably strong in SN\,2023emq and in this regard the SN is more similar to Type Icn SNe. In conclusion, based on our analysis we consider SN\,2023emq a flash-ionised Ibn SN, but the strong emission at $5700$\,Å could indicate that the SN is between the Ibn and Icn populations if it is related to \ion{C}{3}. More spectra obtained immediately after discovery are needed to investigate the importance of the emission at $5700$\,Å and further investigate the diversity of the flash-ionised features in core-collapse SNe.

We also obtained $V$-band polarimetry for SN\,2023emq, making it only the third Type Ibn/Icn SN with optical polarimetry that probed the intrinsic emission from the SN. We find polarisation degree $P=0.55\pm0.30$\% after correction for Galactic ISP and polarisation bias. Based on the upper limit on host extinction, the polarisation is likely not caused by the host dust content, and we conclude that the polarisation is intrinsic to the SN. Given the spectrum is dominated by interaction lines at the time, the result implies that the CSM shows high spherical symmetry in possibly marginally aspherical configuration. More polarimetric observations of these SN classes are required to characterise their geometric diversity and thus gain crucial insight to how the CSM is produced.


\section{Acknowledgments}
We thank the anonymous referee for helpful feedback and Anjasha Gangopadhyay for providing the data of SN\,2019uo.

M.P. and G.L. are supported by a research grant (19054) from VILLUM FONDEN. 
P.C. acknowledges support via an Academy of Finland grant (340613; P.I. R. Kotak).
FEB acknowledges support from ANID-Chile grants BASAL CATA FB210003, FONDECYT Regular 1200495, and Millennium Science Initiative Program  – ICN12\_009. 
T.-W.~Chen thanks the Max Planck Institute for Astrophysics for hosting her as a guest researcher.
L.G. and C.P.G. acknowledges financial support from the Spanish Ministerio de Ciencia e Innovaci\'on (MCIN) and the Agencia Estatal de Investigaci\'on (AEI) 10.13039/501100011033 under the PID2020-115253GA-I00 HOSTFLOWS project, and the program Unidad de Excelencia Mar\'ia de Maeztu CEX2020-001058-M.
L.G. acknowledges support from the European Social Fund (ESF) "Investing in your future" under the 2019 Ram\'on y Cajal program RYC2019-027683-I and from Centro Superior de Investigaciones Cient\'ificas (CSIC) under the PIE project 20215AT01.
C.P.G. acknowledges support from the Secretary of Universities and Research (Government of Catalonia) and by the Horizon 2020 Research and Innovation Programme of the European Union under the Marie Sk\l{}odowska-Curie and the Beatriu de Pin\'os 2021 BP 00168 programme.
J.L. acknowledges support from a UK Research and Innovation Fellowship (MR/T020784/1)
M.N. is supported by the European Research Council (ERC) under the European Union’s Horizon 2020 research and innovation programme (grant agreement No.~948381) and by UK Space Agency Grant No.~ST/Y000692/1.
S.J.S. acknowledges funding from STFC Grant ST/X006506/1 and ST/T000198/1.
P.W. acknowledges support from the Science and Technology Facilities Council (STFC) grant ST/R000506/.

This work is based (in part) on observations collected at the European Organisation for Astronomical Research in the Southern Hemisphere under ESO DDT programme 2111.D-5006 (PI: Pursiainen) and as part of ePESSTO+ under ESO program ID 108.220C (PI: Inserra) and on observations made with the Nordic Optical Telescope, owned in collaboration by the University of Turku and Aarhus University, and operated jointly by Aarhus University, the University of Turku and the University of Oslo, representing Denmark, Finland and Norway, the University of Iceland and Stockholm University at the Observatorio del Roque de los Muchachos, La Palma, Spain, of the Instituto de Astrofisica de Canarias under NOT programmes 67-009. The NOT data presented here were obtained with ALFOSC, which is provided by the Instituto de Astrofisica de Andalucia (IAA) under a joint agreement with the University of Copenhagen and NOT.

This work has made use of data from the Asteroid Terrestrial-impact Last Alert System (ATLAS) project. ATLAS is primarily funded to search for near earth asteroids through NASA grants NN12AR55G, 80NSSC18K0284, and 80NSSC18K1575; by products of the NEO search include images and catalogs from the survey area. The ATLAS science products have been made possible through the contributions of the University of Hawaii Institute for Astronomy, the Queen's University Belfast, the Space Telescope Science Institute, and the South African Astronomical Observatory.  

The Legacy Surveys consist of three individual and complementary projects: the Dark Energy Camera Legacy Survey (DECaLS; Proposal ID \#2014B-0404; PIs: David Schlegel and Arjun Dey), the Beijing-Arizona Sky Survey (BASS; NOAO Prop. ID \#2015A-0801; PIs: Zhou Xu and Xiaohui Fan), and the Mayall z-band Legacy Survey (MzLS; Prop. ID \#2016A-0453; PI: Arjun Dey). DECaLS, BASS and MzLS together include data obtained, respectively, at the Blanco telescope, Cerro Tololo Inter-American Observatory, NSF’s NOIRLab; the Bok telescope, Steward Observatory, University of Arizona; and the Mayall telescope, Kitt Peak National Observatory, NOIRLab. Pipeline processing and analyses of the data were supported by NOIRLab and the Lawrence Berkeley National Laboratory (LBNL). The Legacy Surveys project is honored to be permitted to conduct astronomical research on Iolkam Du’ag (Kitt Peak), a mountain with particular significance to the Tohono O’odham Nation.

NOIRLab is operated by the Association of Universities for Research in Astronomy (AURA) under a cooperative agreement with the National Science Foundation. LBNL is managed by the Regents of the University of California under contract to the U.S. Department of Energy.

This project used data obtained with the Dark Energy Camera (DECam), which was constructed by the Dark Energy Survey (DES) collaboration. Funding for the DES Projects has been provided by the U.S. Department of Energy, the U.S. National Science Foundation, the Ministry of Science and Education of Spain, the Science and Technology Facilities Council of the United Kingdom, the Higher Education Funding Council for England, the National Center for Supercomputing Applications at the University of Illinois at Urbana-Champaign, the Kavli Institute of Cosmological Physics at the University of Chicago, Center for Cosmology and Astro-Particle Physics at the Ohio State University, the Mitchell Institute for Fundamental Physics and Astronomy at Texas A\&M University, Financiadora de Estudos e Projetos, Fundacao Carlos Chagas Filho de Amparo, Financiadora de Estudos e Projetos, Fundacao Carlos Chagas Filho de Amparo a Pesquisa do Estado do Rio de Janeiro, Conselho Nacional de Desenvolvimento Cientifico e Tecnologico and the Ministerio da Ciencia, Tecnologia e Inovacao, the Deutsche Forschungsgemeinschaft and the Collaborating Institutions in the Dark Energy Survey. The Collaborating Institutions are Argonne National Laboratory, the University of California at Santa Cruz, the University of Cambridge, Centro de Investigaciones Energeticas, Medioambientales y Tecnologicas-Madrid, the University of Chicago, University College London, the DES-Brazil Consortium, the University of Edinburgh, the Eidgenossische Technische Hochschule (ETH) Zurich, Fermi National Accelerator Laboratory, the University of Illinois at Urbana-Champaign, the Institut de Ciencies de l’Espai (IEEC/CSIC), the Institut de Fisica d’Altes Energies, Lawrence Berkeley National Laboratory, the Ludwig Maximilians Universitat Munchen and the associated Excellence Cluster Universe, the University of Michigan, NSF’s NOIRLab, the University of Nottingham, the Ohio State University, the University of Pennsylvania, the University of Portsmouth, SLAC National Accelerator Laboratory, Stanford University, the University of Sussex, and Texas A\&M University.

BASS is a key project of the Telescope Access Program (TAP), which has been funded by the National Astronomical Observatories of China, the Chinese Academy of Sciences (the Strategic Priority Research Program “The Emergence of Cosmological Structures” Grant \#XDB09000000), and the Special Fund for Astronomy from the Ministry of Finance. The BASS is also supported by the External Cooperation Program of Chinese Academy of Sciences (Grant \#114A11KYSB20160057), and Chinese National Natural Science Foundation (Grant \#12120101003, \#11433005).

The Legacy Survey team makes use of data products from the Near-Earth Object Wide-field Infrared Survey Explorer (NEOWISE), which is a project of the Jet Propulsion Laboratory/California Institute of Technology. NEOWISE is funded by the National Aeronautics and Space Administration.

The Legacy Surveys imaging of the DESI footprint is supported by the Director, Office of Science, Office of High Energy Physics of the U.S. Department of Energy under Contract No. DE-AC02-05CH1123, by the National Energy Research Scientific Computing Center, a DOE Office of Science User Facility under the same contract; and by the U.S. National Science Foundation, Division of Astronomical Sciences under Contract No. AST-0950945 to NOAO.


%

\vspace{5mm}
\facilities{NOT, NTT, VLT, Swift}


\software{Astropy \citep{AstropyCollaboration2013, AstropyCollaboration2018}, Astro-SCRAPPY \citep{McCully2018}, HOTPANTS \citep{Becker2015},
          Matplotlib \citep{Hunter2007}, Numpy \citep{Harris2020}, SciPy \citep{Virtanen2020}, Source Extractor \citep{Bertin1996}, spalipy \citep{Lyman2021} LMFIT \citep{Newville2014}
          }



\appendix

\section{Bolometric Light Curve}
\label{sec:L_bol}

\begin{figure*}
    \centering
    \begin{minipage}[b]{0.49\textwidth}
            \centering
            \includegraphics[width=\textwidth]{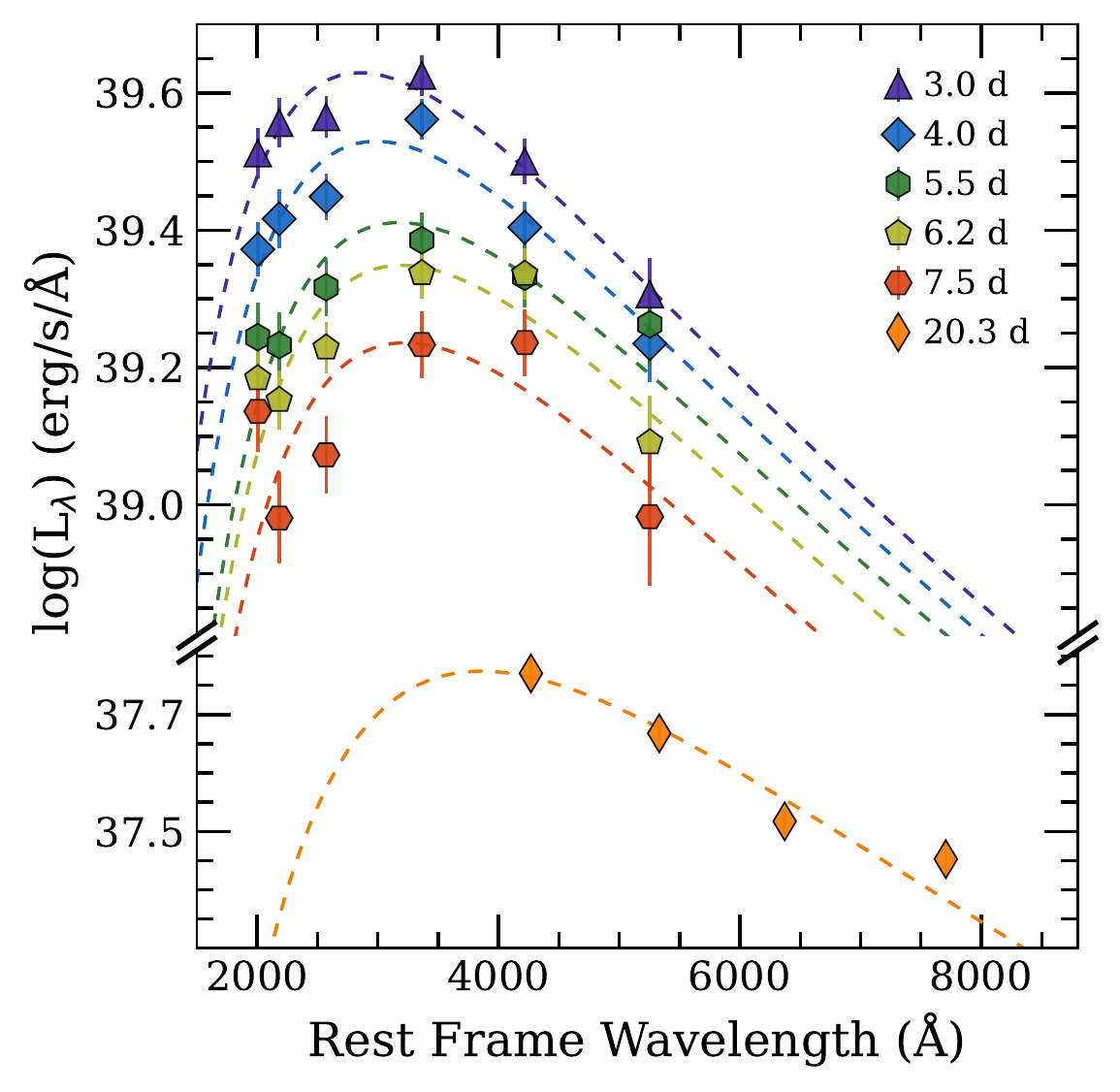}
    \end{minipage} %
    \begin{minipage}[b]{0.49\textwidth}
            \centering
            \includegraphics[width=\textwidth]{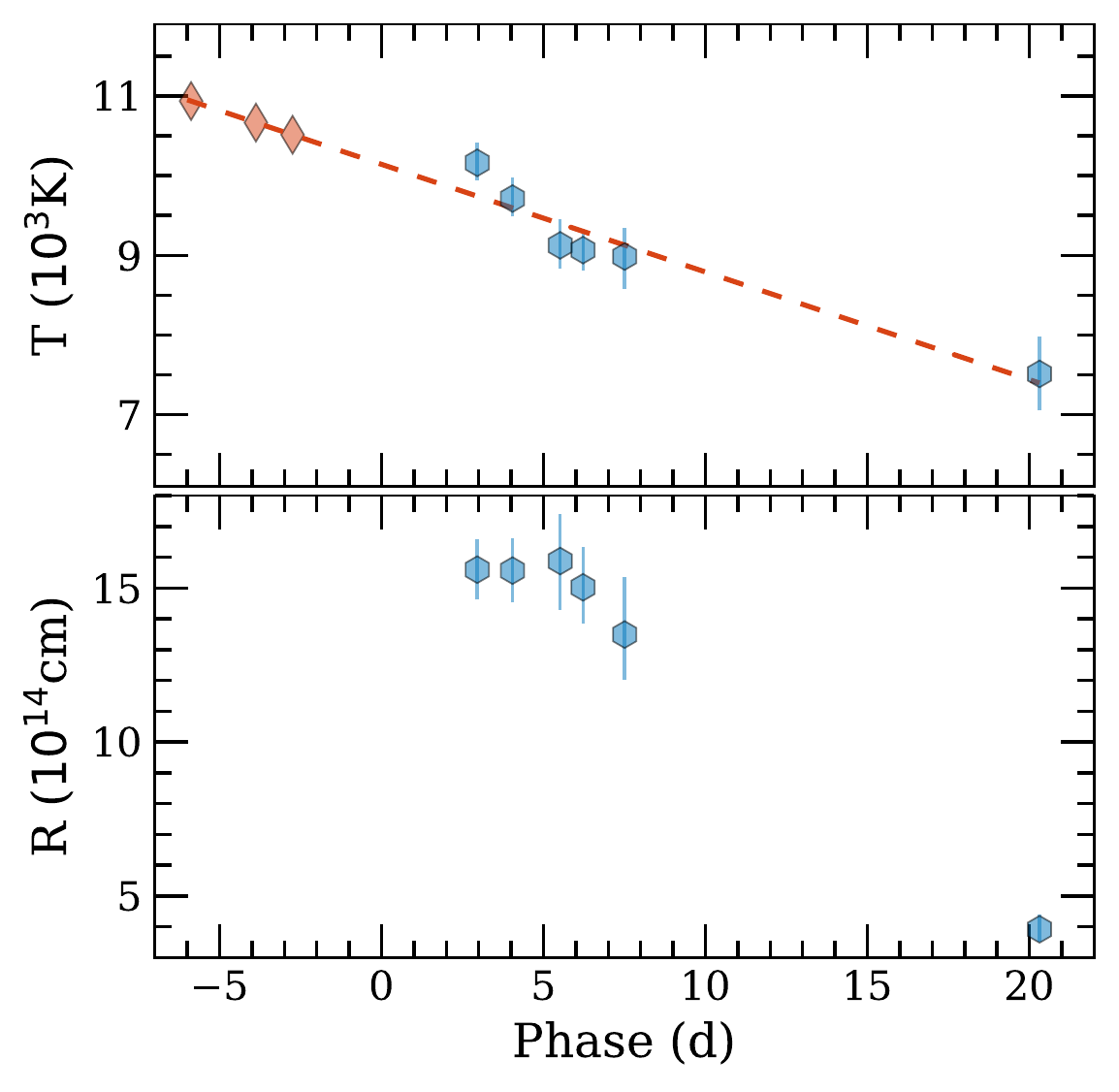}
    \end{minipage} %
    
    \caption{The blackbody modelling of SN\,2023emq. Left: The blackbody fits to the six multi-band epochs. Note the break on the y-axis. Right:  The temperature and radius evolution of SN\,2023emq derived from blackbody fits hexagons). The temperature is linearly declining, and the slope was used to determine the temperature at the time of $o$-band data points on the rise (diamonds).}
    \label{fig:TR_evo}
\end{figure*}

The bolometric light curve of SN\,2023emq was constructed using blackbody fits to the multi-band epochs. The five epochs with UVOT data at $+3$\,--\,8\,d and the NOT $BVRI$ epoch at +20\,d were fit with a blackbody model to determine the evolution of temperature and radius shown in Figure \ref{fig:TR_evo}. For these six epochs, we generated the bolometric luminosity assuming the spectral energy distribution is described by the blackbody fits. During the rise with only $o$-band data, we used the linearly declining temperature curve to estimate the temperature at the time of each data point and estimated the bolometric luminosity by scaling the resulting blackbody to the $o$-band magnitude. We also tested the assumption of constant temperature during the rise by adopting the temperature measured at +3\,d, but the bolometric light curve did not change significantly due to the small change in the temperature. After $+20$\,d, we adopted the temperature found at $+20$\,d ($\sim7500$\,K) and scaled the blackbody to the late-time $R$-band data. As such, our bolometric luminosity is well constrained only between $+3$ and $+20$ days. Outside this range the bolometric luminosities are uncertain and we have assumed 25\% errors.

\section{Tables}

\begin{deluxetable*}{ccccccc}
    \tablecaption{Optical photometry of SN\,2023emq.}
    \tablehead{\colhead{Filter} & \colhead{MJD} & \colhead{Phase (d)} & \colhead{mag} & \colhead{Error} & \colhead{Telescope}}
    \startdata
    $o         $ & $60031.2   $ & $-8.5      $ & $>20.41    $ & $--        $ & ATLAS      \\
    $r         $ & $60032.3   $ & $-7.4      $ & $>19.61    $ & $--        $ & ZTF        \\
    $o         $ & $60033.9   $ & $-5.9      $ & $19.11     $ & $0.14      $ & ATLAS      \\
    $o         $ & $60036.0   $ & $-3.9      $ & $17.68     $ & $0.03      $ & ATLAS      \\
    $o         $ & $60037.2   $ & $-2.7      $ & $17.25     $ & $0.04      $ & ATLAS      \\
    $UVW2      $ & $60043.1   $ & $3.0       $ & $18.32     $ & $0.09      $ & Swift      \\
    $UVM2      $ & $60043.1   $ & $3.0       $ & $18.02     $ & $0.09      $ & Swift      \\
    $UVW1      $ & $60043.1   $ & $3.0       $ & $17.65     $ & $0.08      $ & Swift      \\
    $U         $ & $60043.1   $ & $3.0       $ & $16.91     $ & $0.07      $ & Swift      \\
    $B         $ & $60043.1   $ & $3.0       $ & $16.73     $ & $0.08      $ & Swift      \\
    $V         $ & $60043.1   $ & $3.0       $ & $16.74     $ & $0.13      $ & Swift      \\
    $r         $ & $60043.4   $ & $3.3       $ & $17.44     $ & $0.06      $ & ZTF        \\
    \enddata
    \tablecomments{The magnitudes are corrected for Galactic extinction. NOT/ALFOSC and VLT/FORS2 data are host-subtracted using the last epoch of VLT/FORS2 as template for $R$ band and Legacy Survey data in $g$, $r$ and $i$ bands \citep{Dey2019} for $B$, $V$ and $I$ data, respectively. Only the first few entires are shown here and the complete machine-readable table is available online. (This table is available in its entirety in machine-readable form.)}
    \label{tab:all_phot}
\end{deluxetable*}

\begin{deluxetable*}{cccccccc} 
    \tablecaption{Spectroscopic timeseries of SN\,2023emq. }
   
    \tablehead{\colhead{Date} & \colhead{MJD} & \colhead{Phase (d)} & \colhead{Telescope} & \colhead{Instrument} & \colhead{Grism/Arm} & \colhead{$R$ ($\lambda/\Delta\lambda$)} & \colhead{Range}}
   \startdata
    2023-04-03 &  60038.6 &     -1.3 & FTS   & FLOYDS-S     & --               & 400/700         &     3500 --    10000 \\
    2023-04-14 &  60048.0 &      8.0 & NOT   & ALFOSC       & Gr\#4            & 360             &     3200 --    9600 \\
    2023-04-18 &  60052.1 &     11.6 & NTT   & EFOSC2       & Gr\#11/Gr\#16    & 390/595         &     3380 --    10320 \\
    2023-04-28 &  60062.0 &     21.4 & VLT   & X-Shooter    & UVB/VIS/NIR      & 5400/8900/5600  &     3000 --    24800 \\
    \enddata
    \tablecomments{The first spectrum is the classification one, publicly available in WISeREP.}

 \label{tab:spec_log}
\end{deluxetable*}

\bibliography{bib}{}
\bibliographystyle{aasjournal}



\end{document}